\documentclass[reprint,amsmath,amssymb,aps,onecolumn]{revtex4}
\usepackage{array}
\usepackage{float}
\usepackage{amsmath}
\usepackage{graphicx}

\makeatletter

\providecommand{\tabularnewline}{\\}

\usepackage{dcolumn}% Align table columns on decimal point
\usepackage{bm}% bold math
\newcommand{\op}{\widehat}
\newcommand{\ket}{\rangle}

\newcommand{\+}{\dagger}
\newcommand{\e}{\mathrm{e}}
\newcommand{\dd}{\mathrm{d}}

\makeatother

\begin{document}
\title{Light-matter interaction in open cavities with dielectric stacks}
\author{Astghik Saharyan$^{1}$}
\author{Juan-Rafael \'{A}lvarez$^{2}$}
\author{Thomas H. Doherty$^{2}$}
\author{Axel Kuhn$^{2}$}
\email{axel.kuhn@physics.ox.ac.uk}

\author{St\'{e}phane Guerin$^{1}$}
\email{sguerin@u-bourgogne.fr}

\affiliation{$^{1}$ Laboratoire Interdisciplinaire Carnot de Bourgogne, CNRS UMR
6303, Universit\'{e} Bourgogne Franche-Comt\'{e}, BP 47870, F-21078
Dijon, France.~\\
 $^{2}$ University of Oxford, Clarendon Laboratory, Parks Road, Oxford,
OX1 3PU, UK }
\date{\today}
\begin{abstract}
We evaluate the exact dipole coupling strength between a single emitter
and the radiation field within an optical cavity, taking into account
the effects of multilayer dielectric mirrors. Our model allows one
to freely vary the resonance frequency of the cavity, the frequency
of light or atomic transition addressing it and the design wavelength
of the dielectric mirror. The coupling strength is derived for an
open system with unbound frequency modes. For very short cavities,
the effective length used to determine their mode volume and the lengths
defining their resonances are different, and also found to diverge
appreciably from their geometric length, with the radiation field
being strongest within the dielectric mirror itself. Only for cavities
much longer than their resonant wavelength does the mode volume asymptotically
approach that normally assumed from their geometric length. 
\end{abstract}
\maketitle
%\keywords{Suggested keywords}%Use showkeys class option if keyword
%display desired

%\tableofcontents

\section{Introduction}

The development of universal quantum computation remains a key endeavour
in the coherent control and manipulation of quantum states of light
and matter. A principal component of this effort is improving the
inherent scalability of current architectures; developing methods
for the reliable interconnection of distant qubits and quantum processors \cite{ciracQuantumOpticsWhat2017a}. A promising approach is the
use of high finesse optical cavities to couple individual states of
light and matter, thereby proving a bridge between stationary emitters
and travelling photons as the hosts of quantum information. A strong
coupling within the cavity allows this to be a controllable, deterministic
and inherently reversible interaction, essentially establishing an
idealised quantum interface \cite{ritterElementaryQuantumNetwork2012,kimbleQuantumInternet2008b}.
In principle, this allows for the generation of light-matter entanglement \cite{wilkSingleAtomSinglePhotonQuantum2007}, entanglement swapping \cite{moehringEntanglementSingleatomQuantum2007} and the distribution
of cluster states over an extended quantum network \cite{barrettMultimodeInterferometryEntangling2019}.
In the short term, similar systems have been used extensively for
the enhanced production of single photons, across a variety of emitter
types and platforms such as neutral atoms \cite{Kuhn1999}, ions
\cite{kellerContinuousGenerationSingle2004a}, NV centers \cite{johnsonTunableCavityCoupling2015a}
and quantum dots \cite{dingOnDemandSinglePhotons2016}.

All cavity mediated light-matter coupling schemes developed to date
are based on the Purcell effect, which describes an enhancement to
the spontaneous emission rate of a quantised emitter within a resonant
cavity. When compared to isotropic spontaneous emission into free
space, the rate of emission into the cavity mode is enhanced by the
factor \cite{purcellProceedingsAmericanPhysical1946}: 
\begin{equation}
F_{p}=\dfrac{3\lambda^{3}Q}{4\pi^{2}V}~,\label{eq:PurcellFactor}
\end{equation}
where $Q$ is the quality factor of the resonator, $\lambda$ is the
transition wavelength and $V$ is the optical mode volume of the resonator.
In cavity QED systems which consider coupling to only a single emitter,
this factor is commonly expressed as system cooperativity, given by
$2C=F_{p}=g^{2}/(\kappa\gamma)$, where $g$ is the atom-cavity coupling
rate, $\kappa$ is the cavity field decay rate and $2\gamma$ is the
rate of atomic spontaneous emission. Strongly coupled systems are
correspondingly defined as those where $g\gg(\kappa,\gamma)$, i.e.,
where the coherent coupling rate surpasses incoherent decay mechanisms \cite{kimbleStrongInteractionsSingle1998}. The suppression of photonic
decay within the cavity generally requires the use of highly reflective
mirrors, characterised by a large cavity finesse, $\mathcal{F}=\pi\sqrt{R}/(1-R)$ \cite{saleh91}, where $R$ is the mirror reflectivity. Given the
equivalent representation of finesse as the ratio of cavity free spectral
range to its linewidth, a high finesse additionally ensures good spectral
resolution of each resonance. Therefore, when an emitter with a comparable
transition linewidth is coupled to a high finesse cavity, interaction
with only a single field mode can be assumed.

To achieve the mirror reflectivity required for coherent atom-cavity
interactions, highly reflective dielectric coatings, or Bragg stacks
are used as standard \cite{macleodThinfilmOpticalFilters2018}. These
comprise layer pairs of quarter-wavelength optical thickness dielectric
material, with alternating refractive indices. Generally, a high reflectivity
is only achieved for a large number of such layers, implying a notable
penetration of the stack by incident light. A common strategy for
enhancing cooperativity is the minimisation of mode volume, which
generally requires reducing mirror spacing. At its extreme, the mirror
spacing can be of the same order as the resonant wavelength of interest
\cite{kernEnhancedSinglemoleculeSpectroscopy2014a}. This unintentionally
increases the relative portion of the cavity mode within the dielectric
stack \cite{zhongEffectiveCavityLength1995,macleodThinfilmOpticalFilters2018,apfelOpticalCoatingDesign1977},
rendering the standard optical models of a Fabry-P\'{e}rot resonator
inaccurate. The propagation of light in dielectric stacks has been
studied before under resonance conditions \cite{macleodThinfilmOpticalFilters2018,apfelOpticalCoatingDesign1977,vahalaOpticalMicrocavities2003a}.
Here we go a step further in considering the more general case of
a wave of arbitrary wavelength traveling through the stack. Note that
this is always the case if the emitter coupled to the cavity is not
resonant with the cavity mode.

In order to model the dynamics within a cavity, the quantized light
field is normalized in terms of its quantization volume, which normally
corresponds to the geometric volume of the optical resonator \cite{vogelQuantumOptics2006}.
However, in the case of the ultra-strong coupling resonator we consider
here, modelling the surface of the mirror as a hard boundary to the
cavity mode is no longer appropriate. To rectify this, we consider
an open cavity system, where the electric field is able to propagate
through the dielectric mirror and couples to external free-space modes.
This departs from the standard notion of having a mode volume and
well defined frequency modes of the resonator, suitably modifying
our calculation of the Purcell Factor. Whilst some aspects have already
been addressed in the past, \cite{ujiharaSpontaneousEmissionVery1991,fengTheoryShortOptical1991},
a description which takes into account a large number of dielectric
layers has not been reported, and only the resonant case where the
field has the design wavelength of the dielectric stacks.

In this paper we revise the concept of mode volume and cavity resonance
frequency for a cavity formed from dielectric mirrors. In Section
2 we describe the general quantisation procedure to be used for an
atom within a cavity, where it interacts with a global electromagnetic
field. This procedure departs from the standard input-output formulation
used in the existing literature reporting open optical systems \cite{dutraCavityQuantumElectrodynamics2005,vogelQuantumOptics2006,Walls2008}.
From this, we obtain a general expression for the coupling strength
between the field and atom. As a reasonable approximation, we consider
a cavity which consists of one perfectly reflective mirror and one
partially transparent mirror of finite thickness. At the beginning
of Section 3 we model the partially-transparent cavity mirror as a
single dielectric slab, a case that has been considered extensively
in associated literature \cite{dutraCavityQuantumElectrodynamics2005,vogelQuantumOptics2006}.
Then, we expand it into a multilayer stack, repeating our analysis.
This allows for a discussion on a more realistic specification of
effective cavity length and corresponding mode volume for short cavities.
We show that the expected resonance frequency of the cavity, the resonant
frequency of the emitter and the design frequency of the dielectric
stack may differ substantially from one another. By considering the
boundary to these effects, we demonstrate that the standard models
of optical resonance are asymptotically re-achieved at extended cavity
lengths. Finally, in Section 4 we develop a new model for calculating
the coupling of an emitter to this light field, modelled as an open
quantum system and unconstrained by the surface of the mirrors.

\section{Atom-field coupling\label{sec:Cavity-Atom-coupling}}

The electromagnetic field within a cavity is described by solution
of the Helmholtz equation: 
\begin{align}
\left(\frac{d^{2}}{dx^{2}}+\epsilon_{r}\left(x\right)\frac{\omega^{2}}{c^{2}}\right)\Phi_{\omega}\left(x\right)=0,\label{eq:HelmholtzFreq}
\end{align}
where $\epsilon_{r}\left(x\right)$ is the relative permittivity of
the physical medium, $\Phi_{\omega}\left(x\right)$ is the space-dependent
field eigenmode of continuous index $\omega=2\pi c/\lambda$, where
$\lambda$ is the wavelength of the travelling wave and $c$ is the
speed of light. The orthonormalization condition for these modes is
given by:

\begin{equation}
\left\langle \Phi_{\omega},\Phi_{\omega^{\prime}}\right\rangle =\int_{-\infty}^{\infty}dx\text{ }\epsilon_{r}\left(x\right)\Phi_{\omega}^{*}\left(x\right)\Phi_{\omega^{\prime}}\left(x\right)=\delta\left(\omega-\omega^{\prime}\right).\label{eq:Orthonormality}
\end{equation}
For convenience, eigenmodes can be spatially separated into three
regions: between the mirrors, within the stack, and outside of the
cavity: 
\begin{equation}
\Phi_{\omega}\left(x\right)=\Phi_{\omega,\text{in}}\left(x\right)+\Phi_{\omega,\text{stack}}\left(x\right)+\Phi_{\omega,\text{out}}\left(x\right).\label{eq:ModeDecomposition}
\end{equation}
The corresponding quantized electric field can be written in the Schr\"{o}dinger
picture as: 
\begin{equation}
\hat{E}\left(x\right)=-i\int_{0}^{\infty}d\omega\sqrt{\frac{\hbar\omega}{2\varepsilon_{0}}}\left(\Phi_{\omega}\left(x\right)\hat{a}_{\omega}-\Phi_{\omega}^{*}\left(x\right)\hat{a}_{\omega}^{\dagger}\right),\label{eq:ElectricField0}
\end{equation}
where 
\begin{align*}
\left[\hat{a}_{\omega},\hat{a}_{\omega^{\prime}}^{\dagger}\right] & =\delta\left(\omega-\omega^{\prime}\right),\\
\left[\hat{a}_{\omega},\hat{a}_{\omega^{\prime}}\right] & =0.
\end{align*}
We highlight that the above quantization procedure from first principles
does not break the modes into separate parts (as opposed to any approach
first quantizing the field inside a perfect cavity and subsequently
coupling the field modes to the outside using a variety of input-output
formalisms \cite{Walls2008,dutraCavityQuantumElectrodynamics2005}),
but rather considers the global modes occupying all the Hilbert space.
These global modes behave as uncoupled quantum harmonic oscillators
with a Hamiltonian of the form:

\[
\hat{H}=\int_{0}^{\infty}d\omega\text{ }\hbar\omega\left(\hat{a}_{\omega}^{\dagger}\hat{a}_{\omega}+\frac{1}{2}\right),
\]
where the $1/2$ term is usually removed by renormalizing the zero-point
energy of the field.

To take into account the interaction of the field with an atom localized
between the mirrors, one would like: 
\begin{enumerate}
\item To derive an effective Hamiltonian and master equations accounting
for the dynamics inside of the cavity; 
\item To describe the leakage of the photons from the cavity; 
\item To characterize the photons leaking out of the cavity. 
\end{enumerate}
For which two strategies can be applied: 
\begin{enumerate}
\item Deriving the approximate field Hamiltonians inside and outside the
cavity \textit{before} introducing the atom \cite{dutraCavityQuantumElectrodynamics2005},
which is then coupled to the Hamiltonian within the cavity; 
\item Deriving approximate Hamiltonians where the atom is included in the
cavity initially. 
\end{enumerate}
We focus on the second strategy because we are interested in finding
the most accurate description of emitter-cavity coupling. The presence
of the atom from the outset makes the calculations relatively simpler
to those of \cite{dutraCavityQuantumElectrodynamics2005}, since
the atom, localized in space and interacting with the global modes
$\hat{a}_{\omega}$, leads to a mode-selective behavior of the overall
system.

Let us assume that a point-like two-level atom with states $\vert g\rangle$
and $\vert e\rangle$ separated in energy by $\hbar\omega_{A}$, is
positioned at $x=x_{A}$, between a perfect, mirror placed at $x=-\ell_{c}$
and a partially transparent dielectric mirror positioned at the origin.
Here, $-\ell_{c}<x_{A}<0$, where $\ell_{c}$ is the geometric length
of the cavity. The electric field at the position of the atom can
be written by evaluating Eq. \eqref{eq:ElectricField0} at $x=x_{A}$,
and we consider a field-atom dipolar interaction: 
\[
\hat{V}=-\hat{d}\hat{E}\left(x_{A}\right),
\]
where $\hat{d}=d\hat{\sigma}_{+}+d^{*}\hat{\sigma}_{-}$ is the dipole
moment of the atom, $\hat{\sigma}_{+}=\vert e\rangle\langle g\vert$
and $\hat{\sigma}_{-}=\vert g\rangle\langle e\vert.$ By applying
the rotating wave approximation, the interaction terms read: 
\[
\hat{V}=i\hbar\int_{0}^{+\infty}d\omega\,\left(\eta_{\omega}\hat{\sigma}_{+}\hat{a}_{\omega}-\eta_{\omega}^{*}\hat{a}_{\omega}^{\dagger}\hat{\sigma}_{-}\right),
\]
with a coupling strength given by: 
\begin{align}
\eta_{\omega}=\sqrt{\frac{\omega}{2\hbar\epsilon_{0}}}\,d\,\Phi_{\omega,\text{in}}(x_A).\label{eq:general_coupling}
\end{align}
Therefore, the total Hamiltonian of the atom-cavity global system
is of the form 
\begin{equation}
\begin{split}\hat{H} & =\hbar\omega_{A}\hat{\sigma}_{+}\hat{\sigma}_{-}+\int_{0}^{+\infty}d\omega\,\hbar\omega\,\hat{a}_{\omega}^{\dagger}\hat{a}_{\omega} +i\hbar\int_{0}^{+\infty}d\omega\Bigl(\eta_{\omega}\hat{\sigma}_{+}\hat{a}_{\omega}-\eta_{\omega}^{\ast}\hat{a}_{\omega}^{\dagger}\hat{\sigma}_{-}\Bigr),
\end{split}
\label{eq:Total_Hamiltonin_of_closed_system}
\end{equation}
where the first term corresponds to the energy of the atom, the second
term to the energy of the global electromagnetic field and the third
term the interaction of this field with the atom placed between the
mirrors.

In order to derive an effective Hamiltonian describing the dynamics
of the open atom-cavity system, we need to derive an effective atom-cavity
coupling strength from Eq.~\eqref{eq:general_coupling}. Therefore,
it is necessary to find the modes $\Phi_{\omega}$ that satisfy the
orthonormalization condition Eq.~\eqref{eq:Orthonormality}. We perform
this calculation, taking into account the structure of the partially
transparent mirror of the cavity, in the following section.

\section{Field modes in Bragg stacks\label{sec:FieldModes}}

We first review the propagation of light in a cavity with a thin single-layered
mirror and then build upon it to describe the case of multilayered
stacks. Although this procedure is standard in the literature, it
will prove important to show explicitly the differences that arise
from realistic mirror structures. To ease calculation, we assume that
the electromagnetic field between the mirrors can be separated into
the product of a longitudinal and transverse component \cite{silfvastLaserFundamentals2008}
and only consider the longitudinal propagation in the following.

\subsection{Important parameters}

For the purposes of future reference, we list here the parameters
used to characterise our system in the following sections.

\begin{table}[H]
\begin{centering}
\begin{tabular}{|>{\centering}p{0.7\columnwidth}|>{\centering}p{0.3\columnwidth}|}
\hline 
\textbf{Definition} & \textbf{Notation}\tabularnewline
\hline 
\hline 
Mirror stack design wavelength and angular frequency. & $\lambda_{0},\omega_{0}=\frac{2\pi c}{\lambda_{0}}$\tabularnewline
\hline 
Propagating light wavelength and angular frequency (continuous). & $\lambda,\omega=\frac{2\pi c}{\lambda}$\tabularnewline
\hline 
Geometric cavity length, cavity resonance wavelength and angular frequency. & \tabularnewline
 & $\ell_{c},\lambda_{m},\omega_{m}=\frac{2\pi c}{\lambda_{m}}=\frac{\pi cm}{\ell_{c}}$\tabularnewline
\hline 
Actual cavity resonance angular frequency, leading to a resonance
effective length. & $\omega_{\text{eff}}=\omega^{(m)}_{N}$, $\ell_{\text{eff}}=\frac{\pi cm}{\omega_{\text{eff}}}$\tabularnewline
\hline 
Coupling factor cavity length. & $L^{(m)}_{N}$\tabularnewline
\hline 
\end{tabular}
\par\end{centering}
\caption{Relevant notations for the different wavelengths and frequencies within
the atom-cavity system. The index $m$ corresponds to the number of anti-nodes of the wave between the mirrors, and $N$ to the $2N-1$ dielectric layers.\label{tab:TableNotation}}
\end{table}

\subsection{Single-layered stack \label{subsec:Single-layered-stack}}

\begin{figure}
\begin{centering}
\includegraphics[width=0.7\columnwidth]{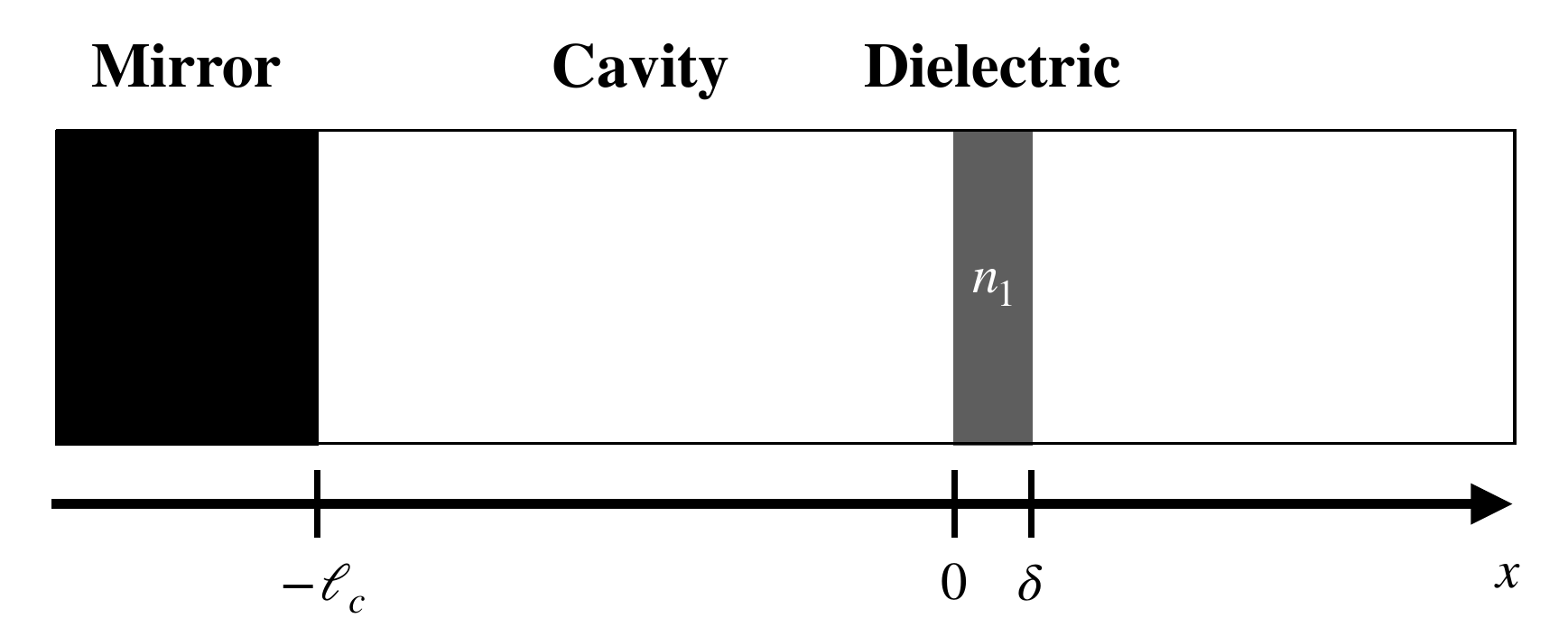}
\par\end{centering}
\caption{Description of the single-layered cavity model. A perfect mirror stands
at $x=-\ell_{c}$, delimiting a cavity of length $\ell_{c}$ with
a partially transparent dielectric material from $x=0$ to $x=\delta$.
\label{fig:SingleLayer}}
\end{figure}

Let us model a cavity of length $\ell_{c}$. For the sake of simplicity,
we assume that such a cavity has a perfect mirror on the left side,
positioned at $x=-\ell_{c}$, and a partially transparent dielectric
placed at the origin, with a refractive index $n_{1}$ and a thickness
$\delta$. The dielectric layer thickness has been designed to support
a target wavelength, $\lambda_{0}$. This structure can be seen in
Fig. \ref{fig:SingleLayer}. In this structure, the spatial distribution
of the electromagnetic field mode $\Phi_{\omega}\left(x\right)$ with
frequency $\omega=2\pi c/\lambda$ is described by the Helmholtz equation
\eqref{eq:HelmholtzFreq}, with a relative permittivity structure
of the form 
\[
\epsilon_{r}\left(x\right)=\begin{cases}
1 & x\in\left[-\ell_{c},0\right]\cup\left[\delta,\infty\right)\\
n_{1}^{2} & x\in\left(0,\delta\right)
\end{cases}.
\]
Under these conditions, Helmholtz's equation is solved by a superposition
of plane waves of the form 
\[
\Phi_{\omega}\left(x\right)=A_{+}e^{i\frac{\omega}{c}\sqrt{\epsilon_{r}\left(x\right)}x}+A_{-}e^{-i\frac{\omega}{c}\sqrt{\epsilon_{r}\left(x\right)}x}.
\]
The distinct solutions are then stitched together by the condition
that $\Phi_{\omega}\left(x\right)$ and its derivative must ensure
continuity throughout the points of discontinuity of $\epsilon_{r}\left(x\right)$,
thus fixing the values of the coefficients $A_{+}$ and $A_{-}$ for
every region. For a single-layered stack, the solution has the form
\begin{equation}
\Phi_{\omega}\left(x\right)=\Phi_{\omega,\text{in}}\left(x\right)\chi_{\left[-\ell_{c},0\right]}+\Phi_{\omega,\text{stack}}\left(x\right)\chi_{\left(0,\delta\right)}+\Phi_{\omega,\text{out}}\left(x\right)\chi_{\left[\delta,\infty\right)},
\end{equation}
following the convention defined by Eq. \eqref{eq:ModeDecomposition}.
Here, $\chi_{\mathcal{D}}$ represents the indicator function within
a domain $\mathcal{D}\subset\mathbb{R}$, i.e., $\chi_{\mathcal{D}}\left(x\right)=1$
for $x\in\mathcal{D}$ and $0$ otherwise. The normalized mode is
described in the three regions by:

\begin{align}
\Phi_{\omega,\text{in}}\left(x\right) & =\frac{2i}{\sqrt{2\pi c{\cal A}}}e^{i\frac{\omega}{c}\ell_{c}}T(\omega)\sin\left[\frac{\omega}{c}(x+\ell_{c})\right],\nonumber \\
\Phi_{\omega,\text{stack}}\left(x\right) & =\frac{1}{\sqrt{2\pi c{\cal A}}}e^{i\frac{\omega}{c}\ell_{c}}\frac{T(\omega)}{1+r_{1}}[(e^{i\frac{\omega}{c}\ell_{c}}-r_{1}e^{-i\frac{\omega}{c}\ell_{c}})e^{i\frac{\omega}{c}n_{1}x}+(r_{1}e^{i\frac{\omega}{c}\ell_{c}}-e^{-i\frac{\omega}{c}\ell_{c}})e^{-i\frac{\omega}{c}n_{1}x}],\label{eq:SingleLayerModes}\\
\Phi_{\omega,\text{out}}\left(x\right) & =\frac{1}{\sqrt{2\pi c{\cal A}}}e^{2i\frac{\omega}{c}\ell_{c}}\dfrac{T(\omega)}{T^{*}(\omega)}e^{i\frac{\omega}{c}x}-e^{-i\frac{\omega}{c}x},\nonumber 
\end{align}
where $\mathcal{A}$ is the transverse area of the mode (a quantity
which is calculated when the mode is normalized in three dimensions.),
and $r$ is the single-layer reflectivity 
\[
r_{1}=\frac{n_{1}-1}{n_{1}+1},
\]
$T(\omega)$ is the cavity spectral response function, which describes the ratio of intensity between the inside and outside of the cavity, with
respect to a particular frequency $\omega$: 
\begin{equation}
T\left(\omega\right)=\frac{t(\omega)}{1+r(\omega)e^{2i\frac{\omega}{c}(\ell_{c}+\frac{\delta}{2})}}.\label{eq:CavityRespFn}
\end{equation}
It depends on the cavity spectral transmission response function $t(\omega)$
\cite{vogelQuantumOptics2006}: 
\[
t(\omega)=\frac{(1-r_{1}^{2})e^{i(n_{1}-1)\frac{\omega}{c}\delta}}{1-e^{2in_{1}\frac{\omega}{c}\delta}r_{1}^{2}},
\]
and spectral reflection response function $r\left(\omega\right)$:
\[
r(\omega)=e^{-i\frac{\omega}{c}\delta}\frac{r_{1}(e^{2in_{1}\frac{\omega}{c}\delta}-1)}{1-e^{2in_{1}\frac{\omega}{c}\delta}r_{1}^{2}}=\left|r\left(\omega\right)\right|e^{i\phi_{r}\left(\omega\right)}.
\]
It is important to stress that both of these parameters are complex
and have associated phases. We will use $\phi_{r}\left(\omega\right)$
in later sections. Together, $t\left(\omega\right)$ and $r\left(\omega\right)$
satisfy the beam splitter relations

\begin{align}
\lvert t(\omega)\rvert^{2}+\lvert r(\omega)\rvert^{2} & =1,\label{eq:one}\\
r^{*}(\omega)t(\omega)+t^{*}(\omega)r(\omega) & =0.\label{eq:zero}
\end{align}

The squared norm of the cavity response function can be decomposed
as a sum of Lorentzian-like functions\cite{vogelQuantumOptics2006,rousseauxControlQuantumTechnologies2016a},
still having $\tilde{\omega}_{m}$ and $\gamma_{1}$ depending on
$\omega$: (see the details in Appendix A):

\begin{equation}
\left|T\left(\omega\right)\right|^{2}=\sum_{m=-\infty}^{\infty}\frac{c}{2L_{1}}\frac{\gamma_{1}\left(\omega\right)}{\left(\omega-\tilde{\omega}_{m}(\omega)\right)^{2}+\left(\frac{\gamma_{1}\left(\omega\right)}{2}\right)^{2}},\label{eq:SingleLayerCREFN}
\end{equation}
where 
\begin{equation}
\begin{split}\gamma_{1}\left(\omega\right) & =-\frac{c}{L_{1}}\ln\left|r\left(\omega\right)\right|,\\
\tilde{\omega}_{m}(\omega) & =m\frac{\pi c}{L_{1}}+\frac{c}{2L_{1}}\left(\pi-\phi_{r}\left(\omega\right)\right),\\
L_{1} & =\ell_{c}+\frac{\delta}{2},
\end{split}
\label{eq:singl_Lorentz_parameters}
\end{equation}
and index 1 in $L_{1}$ and $\gamma_{1}$ indicates that they are
for the case of a single-layer mirror.

In the high-$Q$ limit (reached when assuming a fictitious dielectric
with a high refractive index), and for cavities having $\delta\ll\ell_{c}$,
such that $L_{1}\approx\ell_{c}$, the squared norm of the response
function can be approximated as follows:

\begin{equation}
\left|T\left(\omega\right)\right|^{2}\approx\sum_{m=-\infty}^{\infty}\frac{c}{2\ell_{c}}\frac{\Gamma_{m}}{\left(\omega-\tilde{\omega}_{m}\right)^{2}+\left(\frac{\Gamma_{m}}{2}\right)^{2}},\label{eq:approxSingleLayerCREFN}
\end{equation}
where 
\begin{align}
\Gamma_{m} & =\gamma_{1}(\omega_{m})=-\frac{c}{\ell_{c}}\ln\left|r\left(\omega_{m}\right)\right|,\label{eq:single_width}\\
\tilde{\omega}_{m} & =\omega_{m}+\frac{c}{2\ell_{c}}\left(\pi-\phi_{r}\left(\omega_{m}\right)\right),\label{eq:single_frequency}
\end{align}
with $\omega_{m}=m\pi c/\ell_{c}$. 

At this stage, the terms in the sum in Eq.~\eqref{eq:approxSingleLayerCREFN}
become Lorentzian, with each term centered around the resonance frequency
$\tilde{\omega}_{m}$ and well separated from each other. Taking this
into account we can now write the cavity response function as

\begin{equation}
T\left(\omega\right)\approx\sum_{m=-\infty}^{\infty}\sqrt{\frac{c}{2\ell_{c}}}\frac{\sqrt{\Gamma_{m}}}{\left(\omega-\tilde{\omega}_{m}\right)+i\frac{\Gamma_{m}}{2}}=\sum_{m=-\infty}^{\infty}T_{m}\left(\omega\right),
\end{equation}
where $T_{m}(\omega)$ is so narrow that it does not overlap with
the other Lorentzians $T_{m'}(\omega)$. With this result we can now
write the coupling strength~\eqref{eq:general_coupling} in the form:
\begin{align}
\eta_{\omega}=\sum_{m=-\infty}^{\infty}i\sqrt{\frac{\omega}{\hbar\epsilon_{0}\ell_{c}\mathcal{A}}}d\,e^{i\frac{\omega}{c}\ell_{c}}\sin\big[{\textstyle \frac{\omega}{c}}(x_{A}+\ell_{c})\big]\,\sqrt{\frac{\Gamma_{m}}{2\pi}}\frac{1}{\left(\omega-\tilde{\omega}_{m}\right)+i\frac{\Gamma_{m}}{2}},\label{eq:single_layer_coupling}
\end{align}
which describes the coupling between the global electric field with
the atom localized between the mirrors. The product $\ell_{c}\mathcal{A}$
appearing in the coupling strength can be interpreted as the cavity
mode volume. In the high-Q case considered here, it happens to be
identical to the mode volume determined by the geometric parameters
of the cavity: its mirror separation (or geometric length, $\ell_{c}$),
and mirror area, $\mathcal{A}$. However, we emphasise that the mode
volume was not introduced by normalizing the field in a perfect resonator
of length $\ell_{c}$. Instead, it is a result of the finite width
of the resonances in Eq.~\eqref{eq:approxSingleLayerCREFN}. To this
respect, recovering $\ell_{c}$ has been a coincidence. In many cases
this does not prevail, which shall be demonstrated in the following
discussion.

\subsection{Multilayered stack}

\begin{figure}
\begin{centering}
\includegraphics[width=0.8\columnwidth]{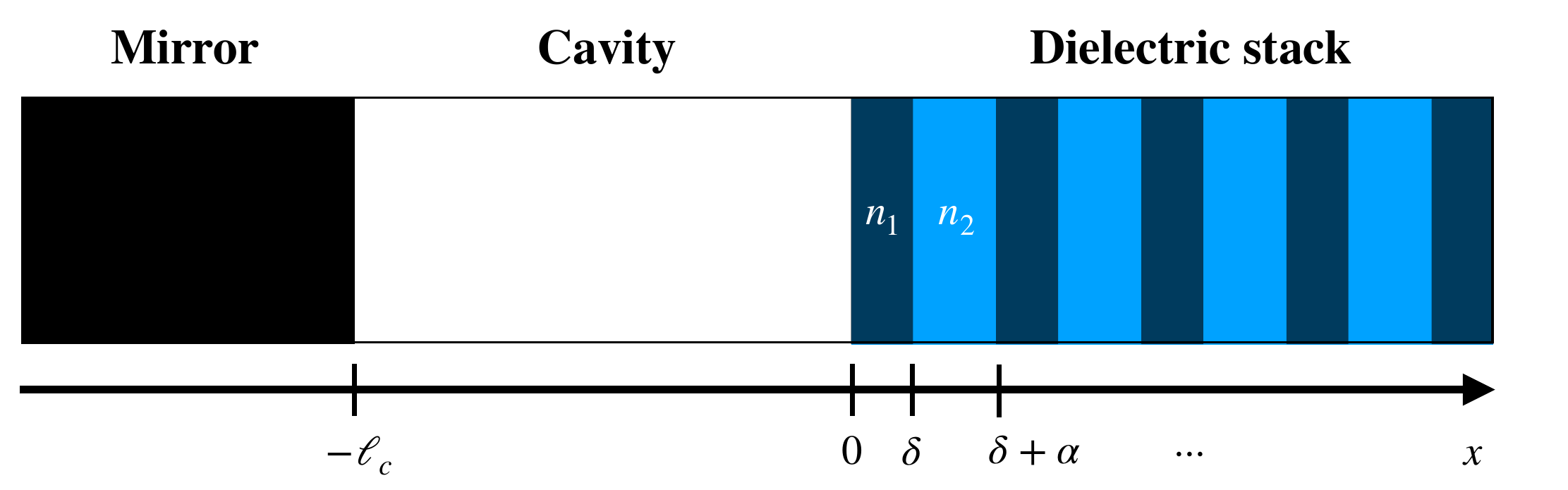}
\par\end{centering}
\caption{Description of the considered model. A perfect mirror stands at $x=-\ell_{c}$,
delimiting a cavity of length $\ell$ with an alternating dielectric
stack standing from zero. \label{fig:CavityModel}}
\end{figure}

We now extend our discussion to consider an alternating stack of two
dielectric materials. The first layer has a width $\delta$ and a
refractive index $n_{1}$, with the other a width $\alpha$ and a
refractive index $n_{2}$. The stack has $2N-1$ layers: $N-1$ pairs
of dielectric and one additional layer of index $n_{1}$. As such,
the first and last layers of the stack have the higher refractive
index ($n_{1}>n_{2}$)\cite{Banning:47}. We do not include in our
model a substrate on which the dielectric coating is deposited, which
would normally correspond to a final, significantly thicker, layer
of low index material. An updated cavity model is shown in Fig. \ref{fig:CavityModel}
and the corresponding relative permittivity is given by: 
\begin{equation}
\epsilon_{r}\left(x\right)=\begin{cases}
1 & -\ell_{c}<x\le0\\
n_{1}^{2} & j_{1}\left(\delta+\alpha\right)<x\leq j_{1}\left(\delta+\alpha\right)+\delta\\
n_{2}^{2} & \left(j_{2}-1\right)\left(\delta+\alpha\right)+\delta<x\leq j_{2}\left(\delta+\alpha\right)\\
1 & x>N\left(\delta+\alpha\right)-\alpha
\end{cases},
\end{equation}
where $j_{1}\in\left\{ 0,1,2,...,N-1\right\} $ and $j_{2}\in\left\{ 1,2,...,N-1\right\} $.
The cavity is designed such that its fundamental longitudinal mode
is at wavelength $\lambda_{0}$, which is also used to specify the
quarter wave dielectric layer optical thickness: $\delta=\lambda_{0}/4n_{1}$
and $\alpha=\lambda_{0}/4n_{2}$. For simplicity, we further consider
$n_{2}=1$. Again, by following the continuity of the field and its
derivative, we solve Helmholtz equation \eqref{eq:HelmholtzFreq}
and find that the mode can be described in terms of the following
functions:

\begin{eqnarray*}
B_{0}(\omega,x) & = & e^{i\frac{\omega}{c}(x+\ell_{c})},\\
B_{j}\left(\omega,x\right) & = & \begin{cases}
\frac{1}{1-r_{1}}\left(B_{j-1}\left(\omega,x_{2}\left(j\right)\right)+r_{1}B_{j-1}^{*}\left(\omega,x_{2}\left(j\right)\right)\right)e^{i\frac{\text{\ensuremath{\omega}}}{c}\left(x-x_{2}\left(j\right)\right)} & j\text{ even},\\
\frac{1}{1+r_{1}}\left(B_{j-1}\left(\omega,x_{1}\left(j\right)\right)-r_{1}B_{j-1}^{*}\left(\omega,x_{1}\left(j\right)\right)\right)e^{i\frac{\omega}{c}n_{1}\left(x-x_{1}\left(j\right)\right)} & j\text{ odd},
\end{cases}
\end{eqnarray*}
where 
\begin{align*}
x_{1}\left(j\right) & =\frac{j-1}{2}\left(\delta+\alpha\right),\\
x_{2}\left(j\right) & =\frac{j}{2}\left(\delta+\alpha\right)-\alpha.
\end{align*}
The solution to the Helmholtz equation is then given by 
\begin{equation}
\Phi_{\omega}\left(x\right)=\sum_{j=0}^{2N}CA_{j}\left(\omega,x\right)\chi_{\Omega_{j}},\label{eq:FullModeDescription}
\end{equation}
where $j$ runs over the $2N-1$ possible layers, labelled as $\Omega_{j}\subset\mathbb{R}$.
The $2N$th term describes the mode exiting the cavity. $C$ is a
normalization factor common to all terms and 
\[
A_{j}\left(\omega,x\right)=B_{j}\left(\omega,x\right)-B_{j}^{*}\left(\omega,x\right).
\]
In particular, $\Omega_{j}$ has the form 
\begin{align*}
\Omega_{0} & =\left[-\ell_{c},0\right],\\
\Omega_{j} & =\begin{cases}
\left[x_{2}\left(j\right),x_{2}\left(j\right)+\alpha\right], & j\text{ }\text{even},\\
\left[x_{1}\left(j\right),x_{1}\left(j\right)+\delta\right], & j\text{ }\text{odd}.
\end{cases}
\end{align*}
Here $A_{0}\left(\omega,x\right)$ corresponds to the mode propagating
between the two mirrors and $A_{2N}\left(\omega,x\right)$ is the
mode corresponding to the outgoing wave.

\begin{figure}
\begin{centering}
\includegraphics[width=1\columnwidth]{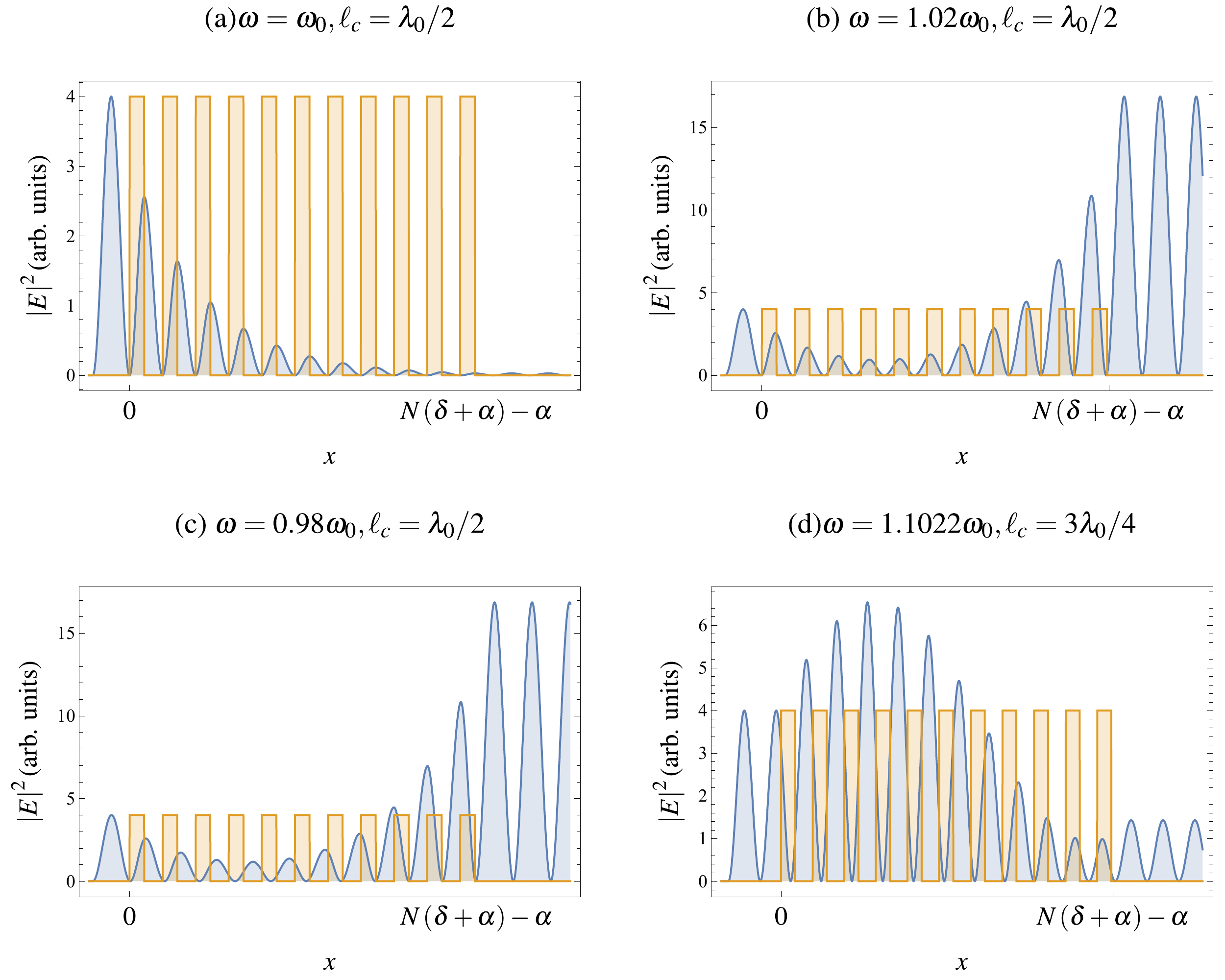}
\par\end{centering}
\caption{Mode propagation of a wave of wavelength $\lambda,$ for a cavity
with a multilayered stack with an index of $n_{1}=1.25$. A perfect
mirror stands at $x=-\ell_{c}$, forming a cavity of length $\ell_{c}$
with an alternating dielectric stack of 21 alternating layers of width
$\delta=\lambda_{0}/4n_{1}$ and $\alpha=\lambda_{0}/4n_{2}$ ($n_{2}$
is taken to be 1). The propagation frequency $\omega=2\pi c/\lambda$
is varied to observe the mode propagation. In (a), the case when $\ell_{c}=\lambda_{0}/2$
and $\omega$ exactly matches with the cavity design resonance frequency
$\omega_{0}$ is shown, obtaining the strongest confinement of light
inside of the cavity. In (b) and (c), we can see how the intensity
changes when the propagating light is slightly off resonance. Finally,
in (d) we increase the cavity length to reach $\ell_{c}=3\lambda_{0}/4$
and observe that we achieve resonance coupling when $\omega=1.1022\omega_{0}$.
Moreover, we can see that in this case the intensity of the light
within the layer stack exceeds the intensity of the light inside of
the cavity. \label{fig:CavityDecay}}
\end{figure}

For a mismatch of the light propagation frequency, $\omega$, with
the stack design frequency, $\omega_{0}=2\pi c/\lambda_{0}$, the
intensity of the mode can either decay or grow exponentially through
the stack. This can potentially be a combination of each, achieving
extremal points of intensity within the dielectric-layer stack. Some
examples are shown in Fig. \ref{fig:CavityDecay}.

\subsection{Effective cavity response function}

After a mode has been described in the form of Eq. \eqref{eq:FullModeDescription},
it is important to consider an effective transmissivity, reflectivity
and response for a multilayered cavity, resembling Eq. \eqref{eq:SingleLayerModes}.
Following the normalization of the modes in \eqref{eq:FullModeDescription},
the description of the inside and outside modes have the form

\begin{align*}
\Phi_{\omega,\text{in}}\left(x\right) & =2ie^{i\frac{\omega}{c}\ell_{c}}\mathcal{T}(\omega)\sin\left[\frac{\omega}{c}(x+\ell_{c})\right],\\
\Phi_{\omega,\text{out}}\left(x\right) & =e^{2i\frac{\omega}{c}\ell_{c}}\dfrac{\mathcal{T}(\omega)}{\mathcal{T}^{*}(\omega)}e^{i\frac{\omega}{c}x}-e^{-i\frac{\omega}{c}x}.
\end{align*}
Here, $\mathcal{T}(\omega)$ is the field amplitude ratio of the multilayered
cavity, which has the form 
\begin{align}
\mathcal{T}(\omega)=\dfrac{e^{-i\frac{\omega}{c}(\ell_{c}+(N-1)(\delta+\alpha))}}{B_{2N-2}^{*}(\omega)}\dfrac{t(\omega)}{1+e^{i\frac{\omega}{c}\delta}e^{i\phi_{B}\left(\omega\right)}r(\omega)},\label{multilayerresp}
\end{align}
where $\phi_{B}(\omega)=\text{\ensuremath{\arg}}\Big(B_{2N-2}/B_{2N-2}^{*}\Big)$
and for simplicity we omitted the argument of $B_{2N-2}=B_{2N-2}(\omega,x_{1}(2N-1))$. Here, $t\left(\omega\right)$ and $r\left(\omega\right)$ remain
the same as for the single layer case.

\begin{figure}[H]
\begin{centering}
\includegraphics[width=0.6\columnwidth]{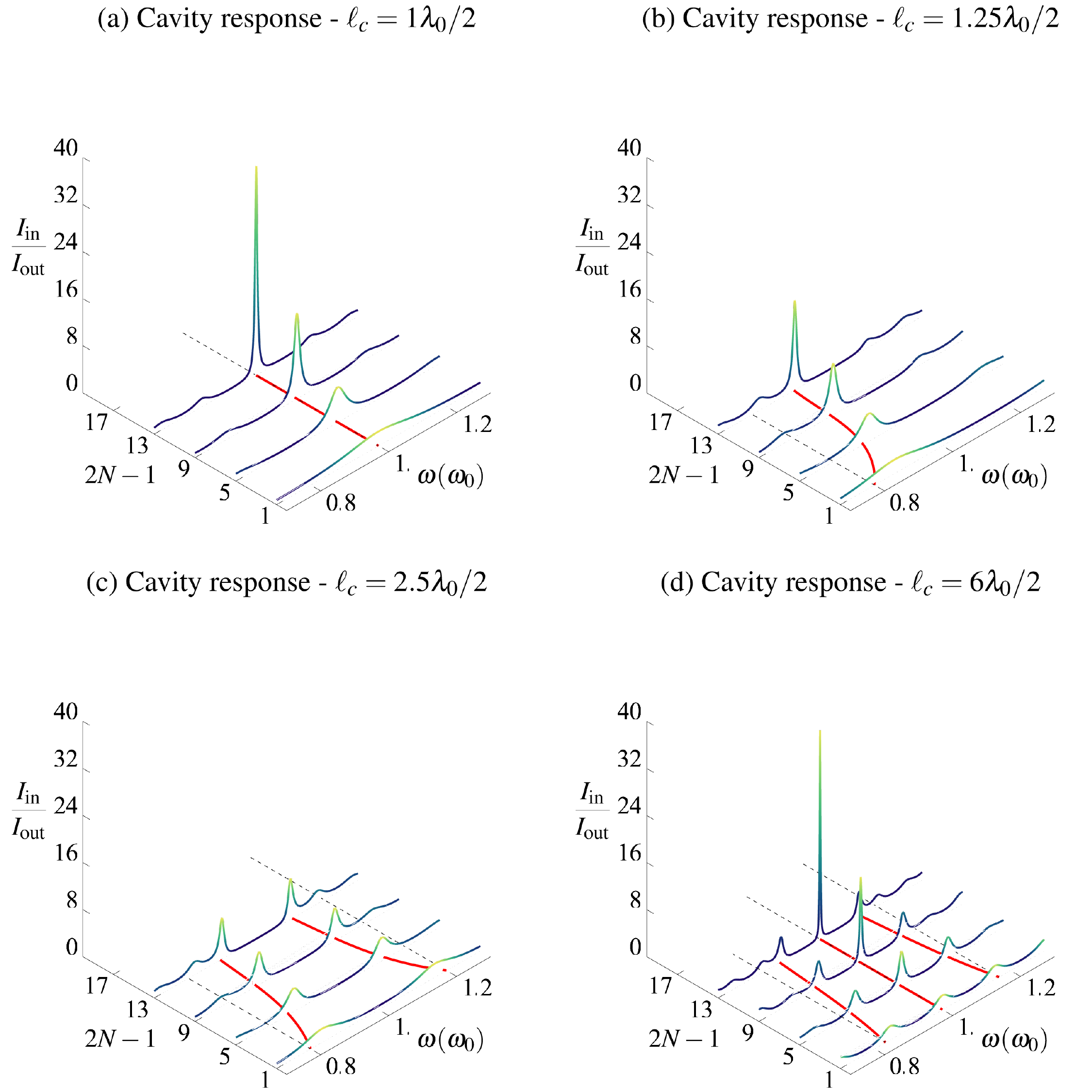}
\par\end{centering}
\caption{Cavity inside-outside intensity ratio, $I_{\text{in}}/I_{\text{out}}$,
as a function of the angular frequency of the light and the number
of layers, $2N-1$, for a cavity with a multilayered stack with a
refractive index $n_{1}=1.25$. A perfect mirror stands at $x=-\ell_{c}$,
and there is an alternating dielectric stack with a varying number
of $N$ layers of width $\delta=\lambda_{0}/4n_{1}$ and $N-1$ layers
of width $\alpha=\lambda_{0}/4n_{2}$ (where $n_{2}$ is taken to
be 1). Both the number of dielectric layers and the spacing of the
cavity are varied to observe different sorts of cavity response functions.
All of these can be decomposed as a sum of Lorentzian functions in
$\omega$. The dashed lines correspond to cavity classical resonances:
$\omega_{m}=m\pi c/\ell_{c}$, where $m$ is the number of antinodes
between the mirrors. The red, solid lines correspond to ${\omega_{\text{eff}}}$,
the peak responses followed by the modes. These do not coincide with
$\omega_{m}$ and are strongly dependent on the number of layers.
\label{fig:CavityResponse}}
\end{figure}

If we introduce an indexing of the response function defined above,
such that $\mathcal{T}(\omega)=\mathcal{T}_{N}(\omega)$, where the label $N$ stands for the $2N-1$ layers, then
it can be shown that 
\[
\dfrac{e^{-i\frac{\omega}{c}(\ell_{c}+(N-1)(\delta+\alpha))}}{B_{2N-2}^{*}(\omega)}=\mathcal{T}_{N-1}(\omega),
\]
where $\mathcal{T}_{N-1}(\omega)$ is the response function of a
cavity having a dielectric stack with $2N-3$ dielectric layers. Fig.
\ref{fig:CavityResponse} shows a series of different cavity response
functions for different numbers of layer stacks and cavity lengths.
Here, our model is seen to exhibit the standard behavior of an optical
cavity by showing a decreased linewidth as the number of layers increases,
as would be expected from increased mirror reflectivity. Further,
the free spectral range of the cavity clearly decreases as its length
is increased.

Just as in the case of the single-layered cavity, we can also write
the square modulus of the response function as a sum of Lorentzian-like
functions, analogous to the one in Eq.~\eqref{eq:SingleLayerCREFN}:

\begin{equation}
|\mathcal{T}(\omega)|^{2}=\sum_{m=-\infty}^{\infty}\frac{c}{\delta\lvert B_{2N-2}(\omega)\rvert^{2}}\frac{\gamma_{N}(\omega)}{\left(\omega-\tilde{\omega}_{m}(\omega)\right)^{2}+\left(\frac{\gamma_{N}(\omega)}{2}\right)^{2}},\label{eq:modsqrtResp}
\end{equation}
where 
\begin{align*}
\gamma_{N}(\omega) & =-c\frac{\ln|r(\omega)|}{\delta/2},\\
\tilde{\omega}_{m} & =\frac{c\pi}{\delta/2}m-c\frac{(\phi_{r}(\omega)+\phi_{B}(\omega)+\pi)}{\delta}.\\
\end{align*}
The term $\phi_{B}(\omega)$ in the expression of $\tilde{\omega}_{m}$
accounts for the multilayer nature of the mirror. In the single-layered
case, this has the simple, analytical form $\phi_{B_{1}}(\omega)=2\ell_{c}\omega/c$,
which leads to the expressions in Eq.~\eqref{eq:singl_Lorentz_parameters}.

As we can see from Fig.~\ref{fig:CavityResponse}, the multilayer
structure leads to a narrowly peaked Lorentzian response function.
In order to obtain the individual Lorentzians corresponding to each
peak in the response function, we apply the following approximation:

\begin{equation}
|\mathcal{T}(\omega)|^{2}\approx\sum_{m}\frac{c}{2L_{N}^{(m)}}\frac{\gamma_{N}^{(m)}}{\left(\omega-\omega^{(m)}_{N}\right)^{2}+\left(\frac{\gamma_{N}^{(m)}}{2}\right)^{2}},\label{eq:MultilayerCREFN-1}
\end{equation}
where $L_{N}^{(m)}$, $\gamma_{N}^{(m)}$ and $\omega^{(m)}_{N}$
are parameters that are found numerically when we fit each individual
peak to the exact cavity response function shown in Eq.~\eqref{eq:modsqrtResp}.
In the case of a single-layered mirror, we obtain the individual Lorentzians
by evaluating the parameters in~\eqref{eq:singl_Lorentz_parameters}
at the resonance frequencies $\tilde{\omega}_{m}$. Here, however,
we cannot follow the same procedure due to the complicated nature
of the coefficient $\lvert B_{2N-2}(\omega)\rvert^{2}$. Therefore,
we apply a numerical fitting to recover the accuracy of the procedure
described above.

\begin{figure}
\begin{centering}
\includegraphics[width=1\columnwidth]{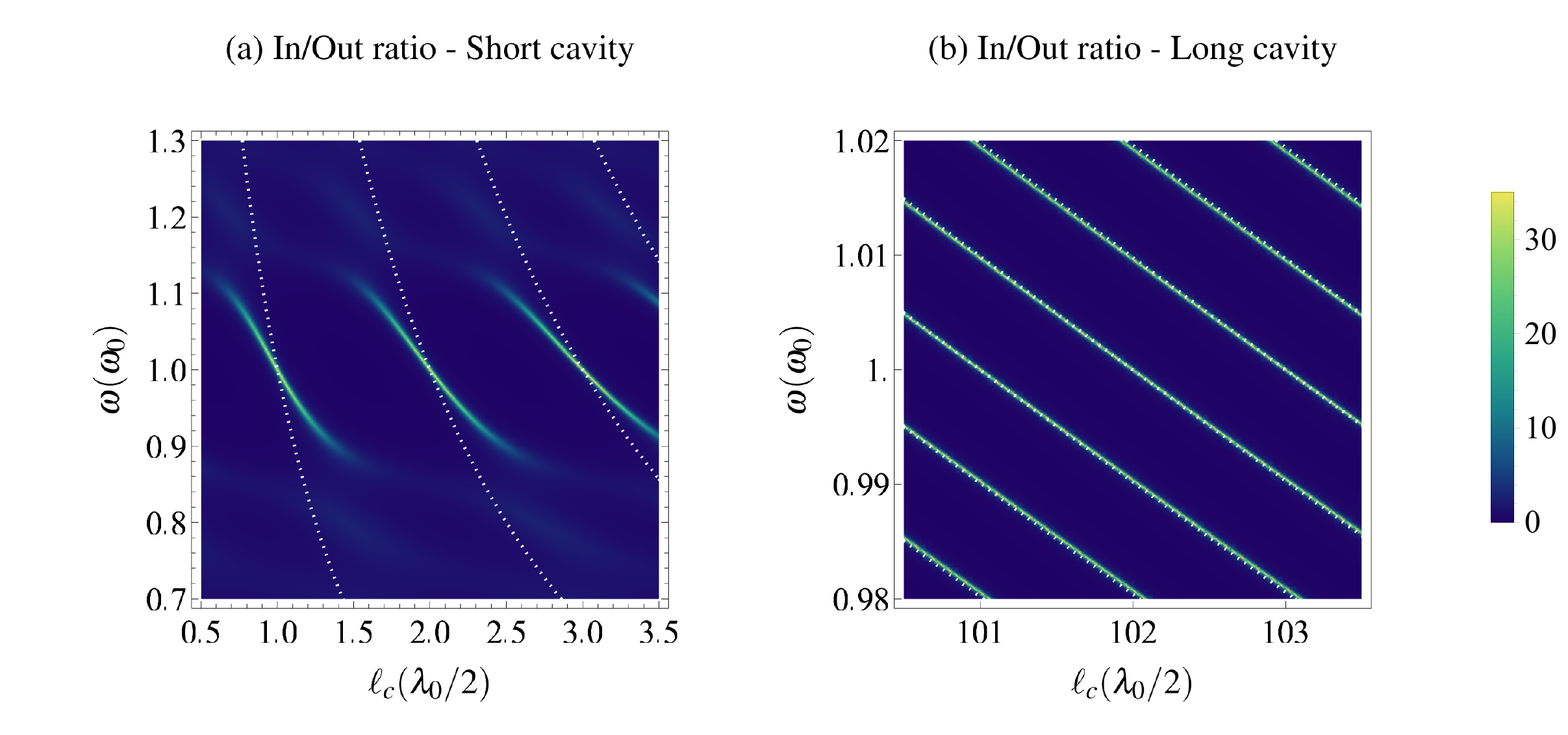}
\par\end{centering}
\caption{Calculated Inside/Outside intensity ratio using the electromagnetic
propagation of the mode versus $\pi mc/\ell_{c}$ shown in white,
dotted lines. When cavities are very short, the mismatch is appreciable
and the maximum inside/outside ratio lines predicted classically differ
significantly with respect to the mode prediction. We can see that,
only at the points where we have a cavity of length $\ell_{c}=p\lambda_{0}/2$,
where $p$ is an integer, the predicted and actual lines match. \label{fig:CavityLength}}
\end{figure}

We will now examine what happens when we vary the spacing between
the mirrors. By taking into account the multilayer structure of the
mirror, we also observe resonance frequency shifts from the expected
resonances, i.e. for a cavity having a mirror spacing $\ell_{c}$,
the expected resonance frequency would be $\omega_{m}=2\pi c/\lambda_{m}=\pi m\left(c/\ell_{c}\right)$,
where $m$ is the number of antinodes between the mirrors. However,
the resonance frequencies that we obtain with a multilayered structure,
in general, do not match the values of $\omega_{m}$ described above
(see Fig. \ref{fig:CavityLength}). In other words, if in the multilayered
case we write $\omega_{\text{eff}}=\omega^{(m)}_{N}=\pi m\left(c/\ell_{\text{eff}}\right)$
for the resonance frequencies, then, in general $\ell_{\text{eff}}$
is different from $\ell_{c}$. In particular, only whenever we have
$\ell_{c}=p\lambda_{0}/2,$ where $p$ is an integer number, then
$\ell_{\text{eff}}$ is the same as $\ell_{c}$. Moreover, the shorter
the cavity, the greater the difference between $\omega_{\text{eff}}$
and $\omega_{m}$.

Each term in Eq. \eqref{eq:MultilayerCREFN-1} corresponds to a well
separated single Lorentzian at a resonance frequency $\omega^{(m)}_{N}$,
hence for the response function we can write

\begin{equation}
\mathcal{T}(\omega)\approx\sum_{m}\sqrt{\frac{c}{2L_{N}^{(m)}}}\frac{\sqrt{\gamma_{N}^{(m)}}}{\left(\omega-\omega^{(m)}_{N}\right)+i\frac{\gamma_{N}^{(m)}}{2}}=\sum_{m}\mathcal{T}_{m}\left(\omega\right),\label{eq:approx_response_function}
\end{equation}
Unlike the single layer case, this expression for the response function
is general; being applicable to cavities of any length, dielectric
layer number and refractive index $n_{1}$. It is now possible to
write the coupling strength \eqref{eq:general_coupling} for this
case: 
\begin{align}
\eta_{\omega}=\sum_{m=-\infty}^{\infty}i\sqrt{\frac{\omega}{\hbar\epsilon_{0}L_{N}^{(m)}\mathcal{A}}}d\,e^{i\frac{\omega}{c}\ell_{c}}\sin\big[{\textstyle \frac{\omega}{c}}(x_{A}+\ell_{c})\big]\,\sqrt{\frac{\gamma_{N}^{(m)}}{2\pi}}\frac{1}{\left(\omega-\omega^{(m)}_{N}\right)+i\frac{\gamma_{N}^{(m)}}{2}} & .\label{eq:MultilayerCoupling}
\end{align}

If we compare this result with the single-layer case of Eq.~\eqref{eq:single_layer_coupling},
we can see that the expressions are the same, with the exception that
for the multilayer case we have $L_{N}^{(m)}$ instead of $\ell_{c}$
and $\gamma_{N}^{(m)}$ instead of $\Gamma_{m}$. Similarly, if
we interpret the product $L_{N}^{(m)}\mathcal{A}$ in the pre-factor
as the mode volume, it is no longer defined through the geometric
length of the cavity, due to the discrepancy between $L_{N}^{(m)}$
and $\ell_{c}$.

Over the width $\gamma_{N}^{(m)}$ of a single Lorentzian, $\omega$
is close to $\omega^{(m)}_{N}$ and varies very slowly, therefore we can
further approximate the mode selective coupling $\eta_{\omega,m}$
as: 
\begin{align}
\eta_{\omega,m}\approx i\sqrt{\frac{\omega^{(m)}_{N}}{\hbar\epsilon_{0}L_{N}^{(m)}\mathcal{A}}}d\,e^{i\frac{\omega}{c}\ell_{c}}\sin\big[{\textstyle \frac{\omega}{c}}(x_{A}+\ell_{c})\big]\,\sqrt{\frac{\gamma_{N}^{(m)}}{2\pi}}\frac{1}{\left(\omega-\omega^{(m)}_{N}\right)+i\frac{\gamma_{N}^{(m)}}{2}} & .
\end{align}

\section{Effective Hamiltonian and Atom-cavity coupling \label{effective_Hamiltonian}}

In section \ref{sec:Cavity-Atom-coupling} we derived a Hamiltonian
describing the global closed system consisting of an atom, cavity
and environment. In this section, our goal is to extract an effective
Hamiltonian to describe the dynamics of the open atom-cavity system.

In further analysis, we limit the Hilbert space to a subspace containing
the excited atom state $\lvert e,0\rangle$ and the single excitation
of the cavity in a superposition of frequency modes: $\lvert g,1\rangle$.
To write the effective Hamiltonian for such a system we define the
following operator: 
\begin{align}
\hat{a}_{m}=\frac{1}{g_{m}}\int_{0}^{\infty}\,d\omega~\eta_{\omega,m}~\hat{a}(\omega),
\label{eq:discrete_modes}
\end{align}
which satisfies the commutation relations: 
\begin{align*}
[\hat{a}_{m},\hat{a}_{m'}^{\dagger}] & =\delta_{mm'},\\{}
[\hat{a}_{m},\hat{a}_{m'}] & =0.
\end{align*}
The definition~\eqref{eq:discrete_modes} allows one to transform the continuous model~\eqref{eq:Total_Hamiltonin_of_closed_system} into discrete one.

It can be shown that
$g_{m}$ is a normalization constant given by  (see Appendix \ref{app:atom_cavity_coupling}): 
\begin{align}
g_{m}=i\sqrt{\frac{\omega^{(m)}_N}{\hbar\epsilon_{0}L_{N}^{(m)}\mathcal{A}}}d\,e^{i\frac{\omega^{(m)}_N}{c}\ell_{c}}\sin\big[{\textstyle \frac{\omega^{(m)}_N}{c}}(x_{A}+\ell_{c})\big]+{\cal O}\left(\epsilon^{2}\right) & ,\label{eq:atom-cavity_resonance_coupling}
\end{align}
up to an error of the order $\epsilon^{2}$, where $\epsilon=\gamma_{N}^{(m)}\frac{x_{A}+\ell_{c}}{c}$.
The width $\gamma_{N}^{(m)}$ decreases with a greater number of
layers and cavity length, hence $\epsilon$ becomes smaller (see Fig.~\ref{fig:CavityResponse})
and the approximation is well validated.

Assuming that only one of the cavity modes at $\omega^{(m)}_N$ is close
to the resonance frequency $\omega_{A}$ of the atom, all other modes
can be safely disregarded and the effective Hamiltonian then reads
(see Appendix~\ref{app:atom_cavity_coupling}): 
\begin{align}
\hat{H}_{\text{eff}} & =\hbar\omega_{A}\hat{\sigma}_{+}\hat{\sigma}_{-}+\left(\hbar\omega^{(m)}_N-i\hbar\frac{\gamma_{N}^{(m)}}{2}\right)\hat{a}_{m}^{\dagger}a_{m}+i\hbar g_{m}\hat{\sigma}_{+}\hat{a}_{m}-i\hbar g_{m}^{\ast}\hat{\sigma}_{-}\hat{a}_{m}^{\dagger},\label{eq:effective_Hamiltonian_on_resonance}
\end{align}
where $\omega^{(m)}_N-\omega_{A}\ll\omega^{(m)}_N$, $g_{m}$ is the coupling
of a single cavity mode $\omega^{(m)}_N$ with the atom and $\gamma_{N}^{(m)}$
is the width of Lorentzian centered at the resonance frequency $\omega^{(m)}_N$.

The expression~\eqref{eq:atom-cavity_resonance_coupling} looks similar
to the atom-cavity coupling for a perfect cavity, i.e., if we write
the perfect cavity Hamiltonian with zero boundary conditions at the
mirrors we obtain: 
\begin{align}
g_{m}=i\sqrt{\frac{\omega_{m}}{\hbar\epsilon_{0}\ell_{c}\mathcal{A}}}d\,e^{i\frac{\omega_{m}}{c}\ell_{c}}\sin\big[{\textstyle \frac{\omega_{m}}{c}}(x_{A}+\ell_{c})\big],
\end{align}
which is different from~\eqref{eq:atom-cavity_resonance_coupling}
by an error factor ${\cal O}\left(\epsilon^{2}\right)$, and again
due to the mismatch between $\ell_{c}$ and $L_{N}^{(m)}$ and $\omega_m$ and $\omega^{(m)}_N$.

For the more general case, when the atom is not necessarily on resonance
with the cavity, (i.e. we consider the states $\lvert e,0\rangle$,
$\lvert g,1\rangle$, as well as $\lvert e,1\rangle$ and $\lvert g,0\rangle$),
it can be shown that the Hamiltonian~\eqref{eq:effective_Hamiltonian_on_resonance}
becomes:

\begin{align}
\hat{H}_{\text{eff}} & =\hbar\omega_{A}\op{\sigma}_{+}\op{\sigma}_{-}+\hbar\sum_{m}\left(\omega^{(m)}_N-i\frac{\gamma^{(m)}_{N}}{2}\right)\op{a}_{m}^{\+}\op{a}_{m}+ig_{m}\op{\sigma}_{+}\op{a}_{m}-ig_{m}^{\ast}\op{\sigma}_{-}\op{a}_{m}^{\+}+i\underline{g}_{m}\op{\sigma}_{-}\op{a}_{m}-i\underline{g}_{m}^{*}\op{\sigma}_{+}\op{a}{\+}_{m},
\end{align}
where 
\begin{align}
\underline{g}_{m}=i\sqrt{\dfrac{\omega^{(m)}_N}{\hbar\epsilon_{0}L_{N}^{(m)}\mathcal{A}}}d^{*}e^{i\frac{\omega^{(m)}_N}{c}\ell_{c}}\sin{\left[\frac{\omega^{(m)}_N}{c}(x_{A}+\ell_{c})\right]}+\mathcal{O}\left(\epsilon^{2}\right).\label{eq:atom-cavity_off_resonance_coupling}
\end{align}

\section{Discussion and Conclusions}

In this paper we have examined light-matter interaction in optical
cavities delimited by multilayered dielectric mirrors using a model
where no assumptions over the longitudinal constraints of the field
have been made. The analysis of the multilayer structure allowed us
to perform a more realistic description of the effective cavity length
and corresponding mode volume.

The effective cavity length $ell_{\text{eff}}$, as defined in the literature
\cite{suematsuDynamicSinglemodeSemiconductor1983,coldrenDiodeLasersPhotonic2012,zhongEffectiveCavityLength1995},
determines the cavity resonance and, as we have shown in this paper,
does not correspond to the coupling factor cavity length $L^{(m)}_N$
used for determining the strength of the cavity-matter coupling. We
have found that these two lengths do not correspond to each other
whenever a multilayer structure is taken into consideration. Additionally,
the resonance frequencies through which $\ell_{\text{eff}}$ is defined
no longer coincide with the expected resonances at integer multiples
of $c/(2l_{c})$. In other words, $\ell_{\text{eff}}$ is different
from the geometric cavity length $\ell_{c}$ unless the latter is
an integer multiple of $\lambda_{0}/2$, where $\lambda_{0}$ is the
stack design wavelength. A pronounced side effect of this discrepancy
is that the most common approach \cite{saleh91} for determining
a cavity length by means of measuring its free-spectral range is bound
to fail for very short cavities.The geometric length of the cavity
might differ by up to $\lambda_{0}/2$ from its effective length determining
the free spectral range. This difference may be of importance in applications
where the cavity spacings used are very short and the mirrors are
made of dielectric coatings \cite{trichetNanoparticleTrappingCharacterization2016,maderScanningCavityMicroscope2015}.

The standard model for light-matter interaction in optical cavities
considers the mode volume of the cavity to be the product between
the longitudinal extent of the mode (considered to be only the separation
between the mirrors) and a factor accounting for the transverse behavior
of the mode \cite{silfvastLaserFundamentals2008}. However, here
we have shown that this is only valid if the mirrors can be seen as
hard mode-delimiting boundaries. Any physical cavity is very different
to that respect. The mode penetrates into the dielectric mirror stack,
couples to the outside and the light frequency is not quantized in
these open systems. However, even though we don't explicitly consider
a mode volume, we show that the atom-cavity coupling involves a factor
having the unit of volume. Only in the case of a perfect cavity having
hard boundaries at the positions of the mirrors, this factor coincides
with the geometric definition of the mode volume. In the multilayered
case, however, this factor does not necessarily correspond to the
geometric volume of the cavity, particularly due to the discrepency
between the geometric length $\ell_{c}$, the effective length $\ell_{\text{eff}}$
and $L^{(m)}_N$. More specifically, for a cavity with a mirror spacing $\ell_c=\lambda_0/2$, the effective cavity length $\ell_{\text{eff}}$ coincides with $\ell_c$, however the coupling factor cavity length does not: $L^{(1)}_{19}\approx3.06\ell_c$ . If we increase the mirror spacing, such that $\ell_c=1.8\lambda_0/2$, then for the other parameters we obtain:  $\ell_{\text{eff}}\approx0.53\ell_c$ and $L^{(1)}_{19}\approx2.4\ell_c$. As expected, for longer cavities the discrepancy between these terms decreases, e.g., when $\ell_c=30.8\lambda_0/2$ we obtain $\ell_{\text{eff}}\approx0.97\ell_c$ and $L^{(q)}_{19}\approx1.09\ell_c$, where $q$ is the number of the mode with highest amplitude. This discrepancy between the coupling factor cavity length $L^{(m)}_{N}$  and geometric cavity length $\ell_{c}$ can lead to a Purcell factor different
from the one calculated by using a standard approach, especially with
short mirror spacings.

Eventually, from the general quantized field we extract an effective
Hamiltonian describing an open cavity system. From this we explicitly
calculate the atom-cavity coupling rate both for on and off resonant
behavior. This quantization from first principles of more realistic
open cavities is a preliminary step for considering the complex dynamics
featuring a laser-driven atom in such cavities. Work is ongoing to
derive microscopic models \cite{PhysRevA.101.012122} for controlling
photonic states produced from such cavities \cite{Kuhn1999}.

We also note that submicrometric confinement of light in (dissipative
and dispersive) metallic media gives rise to surface plasmon polaritons
that can be used for quantum optics at nanoscale \cite{changSinglephotonTransistorUsing2007}.
Albeit these are no multilayered structures, the construction of quantized
models taking into account the resulting losses follows a similar
approach and is based on a microscopic oscillator model for the medium
coupled to the electromagnetic field\cite{huttnerQuantizationElectromagneticField1992}
with a particular care for finite-size media \cite{dorierCanonicalQuantizationQuantum2019,dorierCriticalReviewQuantum2020}.
\begin{acknowledgments}
J.R.A. acknowledges Christian Poveda regarding code optimization.
All authors acknowledge Mark IJspeert, Marwan Mohammed and Ezra Kassa
for useful discussions. 
\end{acknowledgments}

\section*{Funding}

European Union Horizon 2020 (Marie Sklodowska-Curie 765075-LIMQUET).
\\
 EPSRC through the quantum technologies programme (NQIT hub, EP/M013243/1).
\\
 EIPHI Graduate School (ANR-17-EURE-0002).

\bibliography{Paper}
% Produces the bibliography via BibTeX.

\newpage

\appendix
%dummy comment inserted by tex2lyx to ensure that this paragraph is not empty

\section{Lorentzian structure of the cavity spectral response function}

\label{app:Lorentzian_decomposition}

In this Appendix we derive the Lorentzian structure of the cavity
response function~\eqref{eq:CavityRespFn}~\cite{dutraCavityQuantumElectrodynamics2005}:
\begin{equation}
T\left(\omega\right)=\frac{t(\omega)}{1+r(\omega)e^{2i\frac{\omega}{c}L_{1}}}.
\end{equation}
Writing the square modulus of $T\left(\omega\right)$, and using conditions~\eqref{eq:one}
and \eqref{eq:zero} we get 
\begin{align*}
|T(\omega)|^{2} & =\frac{1-|r|^{2}}{|1+|r|e^{i\Phi}|^{2}}=\Big[1-\frac{|r|e^{i\Phi}}{1+|r|e^{i\Phi}}-\frac{|r|e^{-i\Phi}}{1+|r|e^{-i\Phi}}\Big],
\end{align*}
where for simplicity we have not written the dependence of $r(\omega)$
on $\omega$, and we have used the following notations $r(\omega)=|r|e^{i\Phi_{r}(\omega)}$,
$\Phi=2\frac{\omega}{c}L_{1}+\Phi_{r}(\omega)$. Using the geometric series
formula 
\begin{align*}
\sum_{n=1}^{+\infty}~q^{n}=\frac{q}{1-q},\quad|q|<1
\end{align*}
we can write the above expression for the response function as follows
\begin{align*}
|T(\omega)|^{2} & =1+\sum_{n=1}^{+\infty}~|r|^{n}\Big(e^{in(\Phi+\pi)}+e^{-in(\Phi+\pi)}\Big)\\
 & =\sum_{n=-\infty}^{+\infty}~|r|^{|n|}e^{in(\Phi+\pi)}.
\end{align*}
We further apply the Poisson summation formula, which states 
\begin{align*}
\sum_{n=-\infty}^{+\infty}f(n) & =\sum_{m=-\infty}^{+\infty}\int_{-\infty}^{+\infty}dx~f(x)e^{-i2\pi mx}=\sum_{m=-\infty}^{+\infty}\tilde{f}(m)
\end{align*}
where $\tilde{f}(m)=\mathbf{F}_{m}\big[f(x)\big]$ is the Fourier
transform of $f(x)$. We apply the Fourier transform to the function
\begin{align*}
f(x) & =|r|^{|x|}e^{ix(\Phi+\pi)}=e^{|x|\ln|r|+ix(\Phi+\pi)}
\end{align*}
and obtain 
\begin{align*}
\mathbf{F}_{m}\big[f(x)\big]=\frac{2a}{a^{2}+\alpha^{2}},
\end{align*}
where $a=-\ln|r|$, $\alpha=\Phi+\pi-2\pi m$. Hence, 
\begin{align*}
\tilde{f}(m) & =\frac{-2\ln|r|}{(\ln|r|)^{2}+((\Phi+\pi)-2\pi m)^{2}}.
\end{align*}
We further expand this expression by writing the explicit expression
for $\Phi$: 
\begin{align*}
\tilde{f}(m) & =\frac{-2\ln|r|}{(\ln|r|)^{2}+((2\frac{\omega}{c}L_{1}+\Phi_{r}+\pi)-2\pi m)^{2}},
\end{align*}
and multiplying both the numerator and the denominator by $(c/2L_{1})^{2}$
we find the Lorentzian structure of the cavity spectral response function,
still having the parameters $\tilde{\omega}_{m}$ and $\gamma_1$ depend
on $\omega$: 
\begin{align*}
|T(\omega)|^{2} & =\sum_{n=-\infty}^{+\infty}~\frac{c}{2L_{1}}\frac{\gamma_1(\omega)}{(\omega-\tilde{\omega}_{m}(\omega))^{2}+\Big(\frac{\gamma_1(\omega)}{2}\Big)^{2}},\\
\gamma_1(\omega) & =-\frac{c}{L_{1}}\ln|r(\omega)|,\\
\tilde{\omega}_{m} & =\frac{\pi c}{L_{1}}m-\frac{c}{2L_{1}}(\Phi_{r}(\omega)+\pi),
\end{align*}

In the limit of large refractive indices, the reflection coefficient
becomes $r(\omega)=r_{1}e^{-i(\pi+\frac{\omega}{c}\delta)}$, i.e
$\Phi_{r}(\omega)=-(\pi+\frac{\omega}{c}\delta)$, which tends to
$-\pi$ in the limit of a negligible width $\delta$ of the dielectric,
and we recover the expression for a perfect cavity.

\section{Effective Hamiltonian}

\label{app:atom_cavity_coupling}

In order to extract the description of the atom-cavity open system
from the one described in section \ref{sec:Cavity-Atom-coupling},
we first solve the Schr\"{o}dinger equation

\begin{align}
i\hbar\dfrac{\partial}{\partial t}\left|\psi\left(t\right)\right\rangle =\hat{H}\left|\psi\left(t\right)\right\rangle .\label{eq:Shrodinger_equation}
\end{align}
If we put the Hamiltonian \eqref{eq:Total_Hamiltonin_of_closed_system}
and the following wavefunction 
\begin{align}
\left|\psi\left(t\right)\right\rangle =\int_{0}^{+\infty}d\omega e^{-i\omega t}c_{g,1}(\omega,t)\left|g,1\right\rangle +e^{-i\omega_{A}t}c_{e,0}(t)\left|e,0\right\rangle \label{wave_function_Schrod_repres}
\end{align}
into Eq.~\eqref{eq:Shrodinger_equation}, we obtain the dynamical
equations for the coefficients $c_{g,1}(\omega,t)$, $c_{e,0}(t)$ \cite{rousseauxControlQuantumTechnologies2016a}:

\begin{align}
\dot{c}_{g,1}(\omega,t) & =-\sum_{m}\eta_{\omega,m}^{*}e^{i(\omega-\omega_{A})t}c_{e,0}(t)\label{eq:B3a}\\
\dot{c}_{e,0}(t) & =\int_{0}^{+\infty}d\omega\,\sum_{m}\eta_{\omega,m}e^{-i(\omega-\omega_{A})t}c_{g,1}(\omega,t)
\end{align}
In order to trace out the continuous degrees of freedom from the dynamics,
we define the integrated, mode-selective probability amplitude: 
\begin{align}
c_{g,1}^{(m)}(t) & =\dfrac{1}{g_{m}}\int_{0}^{+\infty}d\omega\,\eta_{\omega,m}e^{-i(\omega-\omega_{A})t}c_{g,1}(\omega,t).\label{mode_select}
\end{align}
We now write the time derivative of this quantity, using the property
\begin{align*}
\eta_{\omega,m}^{*}\eta_{\omega,m'}=\delta_{mm'}\lvert\eta_{\omega,m}\rvert^{2},
\end{align*}
which is a consequence of the fact that the response function is a
sum of narrow peaked and well separated Lorentzians. Using this property
we get the equation of motion for the $m$-th mode, using the integration
of \eqref{eq:B3a} and inverting the order of time and frequency integration:
\begin{equation}
\begin{split}\dot{c}_{g,1}^{(m)}(t) & =\dot{c}_{g,1}^{(m,0)}(t)-\dfrac{1}{g_{m}}\int_{0}^{+\infty}d\omega\,\lvert\eta_{\omega,m}\rvert^{2}c_{e,0}(t)\\
 & +\dfrac{i}{g_{m}}\int_{0}^{t}dt'c_{e,0}(t')\int_{0}^{+\infty}d\omega\,\lvert\eta_{\omega,m}\rvert^{2}(\omega-\omega_{A})e^{-i(\omega-\omega_{A})(t-t')}
\end{split}
\label{eq:time_derivative_of_mode_select}
\end{equation}
where we introduce the time derivative of the initial time defined
as follows: 
\begin{align*}
c_{g,1}^{(m,0)}=\dfrac{1}{g_{m}}\int_{0}^{+\infty}d\omega\,\eta_{\omega,m}e^{-i(\omega-\omega_{A})t}c_{g,1}(\omega,0).
\end{align*}
Two integrals appear in ~\eqref{eq:time_derivative_of_mode_select}:
\begin{align*}
\mathcal{I}_{1} & =\int_{0}^{+\infty}d\omega\,\lvert\eta_{\omega,m}\rvert^{2},\\
\mathcal{I}_{2} & =\int_{0}^{+\infty}d\omega\,\lvert\eta_{\omega,m}\rvert^{2}(\omega-\omega_{A})e^{-i(\omega-\omega_{A})(t-t')},
\end{align*}
which can be evaluated using complex contour methods, and in the limit
of $\epsilon=\gamma_N^{(m)}\frac{x_{A}+\ell}{c}\ll1$, we obtain (here
after in the parameters $\gamma^{(m)}_{N}$, $L^{(m)}_{N}$ and $\omega^{(m)}_{N}$ we omit the index indicating
the number of layers) 
\begin{align*}
\mathcal{I}_{1} & =\dfrac{\omega^{(m)}|d|^{2}}{\hbar\epsilon_{0}\mathcal{A}L^{(m)}}\sin^{2}{\left[2\frac{\omega^{(m)}}{c}(x_{A}+\ell_{c})\right]},\\
\mathcal{I}_{2} & =\frac{\omega^{(m)}\lvert d\rvert^{2}}{\hbar\epsilon_{0}\mathcal{A}L^{(m)}}\sin^{2}{\left[2\frac{\omega^{(m)}}{c}(x_{A}+\ell_{c})\right]}\left(\Delta_{m}-i\dfrac{\gamma^{(m)}}{2}\right)e^{-i(\Delta_{m}-i\frac{\gamma^{(m)}}{2})(t-t')},
\end{align*}
where $\Delta_{m}=\omega^{(m)}-\omega_{A}$. We note that, since we
defined $g_{m}$ as the normalization coefficient of \eqref{mode_select},
then the integrals $\mathcal{I}_{1}$ and $\mathcal{I}_{2}$ become
\begin{align*}
\mathcal{I}_{1} & =\lvert g_{m}\rvert^{2},\\
\mathcal{I}_{2} & =\lvert g_{m}\rvert^{2}\left(\Delta_{m}-i\dfrac{\gamma^{(m)}}{2}\right)e^{-i(\Delta_{m}-i\frac{\gamma^{(m)}}{2})(t-t')},
\end{align*}
with 
\begin{align*}
g_{m} & =i\sqrt{\dfrac{\omega^{(m)}}{\hbar\epsilon_{0}\mathcal{A}L^{(m)}}}d\e^{i\frac{\omega^{(m)}}{c}\ell_{c}}\sin{\left[\frac{\omega^{(m)}}{c}(x_{A}+\ell_{c})\right]}+\mathcal{O}\left(\epsilon^{2}\right).
\end{align*}

Having obtained the expression for $g_{m}$, we can now rewrite the
dynamical equation \eqref{eq:time_derivative_of_mode_select} as follows:
\begin{equation}
\label{c_g_1_m_dynamics}
\begin{split}
\dot{c}_{g,1}^{(m)}=\dot{c}_{g,1}^{(m,0)} & -g_{m}^{*}c_{e,0}(t) \\
 & +ig_{m}^{*}\int_{0}^{t}dt'c_{e,0}(t')\left(\Delta_{m}-i\dfrac{\gamma^{(m)}}{2}\right)\e^{-i(\Delta_{m}-i\frac{\gamma^{(m)}}{2})(t-t')}
 \end{split}
\end{equation}
If we formally integrate equation~\eqref{eq:B3a} and put the result
in \eqref{mode_select}, it can be shown that 
\begin{align*}
c_{g,1}^{(m)}(t)-c_{g,1}^{(m,0)}=-g_{m}^{*}\int_{0}^{t}dt'c_{e,0}(t')e^{-i(\Delta_{m}-i\frac{\gamma^{(m)}}{2})(t-t')},
\end{align*}
hence Eq.~\eqref{c_g_1_m_dynamics} becomes 
\begin{align*}
\dot{c}_{g,1}^{(m)}=\dot{c}_{g,1}^{(m,0)}(t)-g_{m}^{*}c_{e,0}(t)-i\left(\Delta_{m}-i\frac{\gamma^{(m)}}{2}\right)(c_{g,1}^{(m)}(t)-c_{g,1}^{(m,0)}(t)).
\end{align*}
Considering that at initial time $t=0$ the atom is in its excited
state and there is no photon in the cavity, we finally get the following
set of dynamical equations 
\begin{equation}
\begin{split}\dot{c}_{g,1}^{(m)} & =-g_{m}^{*}c_{e,0}(t)-i\left(\Delta_{m}-i\frac{\gamma^{(m)}}{2}\right)c_{g,1}^{(m)}\\
\dot{c}_{e,0} & =\sum_{m}g_{m}c_{g,1}^{(m)}(t),
\end{split}
\end{equation}
which then yields that for the state described by the wavefunction
\begin{align*}
\lvert\psi(t)\ket=\sum_{m}c_{g,1}^{(m)}\lvert g,1_{\omega}\ket+c_{e,0}(t)\lvert e,0\ket
\end{align*}
the Hamiltonian can be written as 
\begin{align*}
\hat{H}_{\text{eff}} & =\hbar\sum_{m}\left(\Delta_{m}-i\frac{\gamma^{(m)}}{2}\right)\op{a}_{m}^{\+}\op{a}_{m}+ig_{m}\op{\sigma}_{+}\op{a}_{m}-ig_{m}^{\ast}\op{\sigma}_{-}\op{a}_{m}^{\+}.
\end{align*}
Finally, we can write the effective Hamiltonian back from the rotating
frame, via the transformation: 
\begin{align*}
R=e^{i\omega_{A}t(\op{\sigma}_{+}\op{\sigma}_{-}+\sum\op{a}_{m}^{\+}\op{a}_{m})},
\end{align*}
leading to an effective Hamiltonian describing the dynamics restricted
between the mirrors: 
\begin{align}
\hat{H}_{\text{eff}} & =\hbar\omega_{A}\op{\sigma}_{+}\op{\sigma}_{-}+\hbar\sum_{m}\left(\omega^{(m)}-i\frac{\gamma^{(m)}}{2}\right)\op{a}_{m}^{\+}\op{a}_{m}+ig_{m}\op{\sigma}_{+}\op{a}_{m}-ig_{m}^{\ast}\op{\sigma}_{-}\op{a}_{m}^{\+}.
\end{align}

It can be shown that in the case where we have off-resonant atom-cavity
coupling, the effective Hamiltonian can be written as follows
\begin{align}
\hat{H}_{\text{eff}} & =\hbar\omega_{A}\op{\sigma}_{+}\op{\sigma}_{-}+\hbar\sum_{m}\left(\omega^{(m)}-i\frac{\gamma^{(m)}}{2}\right)\op{a}_{m}^{\+}\op{a}_{m}+ig_{m}\op{\sigma}_{+}\op{a}_{m}-ig_{m}^{\ast}\op{\sigma}_{-}\op{a}_{m}^{\+}+i\underline{g}_{m}\op{\sigma}_{-}\op{a}_{m}-i\underline{g}_{m}^{*}\op{\sigma}_{+}\op{a}{\+}_{m},
\end{align}
where 
\begin{align}
\underline{g}_{m}=i\sqrt{\dfrac{\omega^{(m)}}{\hbar\epsilon_{0}L^{(m)}\mathcal{A}}}d^{*}e^{i\frac{\omega^{(m)}}{c}\ell_{c}}\sin{\left[\frac{\omega^{(m)}}{c}(x_{A}+\ell_{c})\right]}+\mathcal{O}\left(\epsilon^{2}\right).
\end{align}

%\bibliography{Paper}

% The \nocite command causes all entries in a bibliography to be printed out
% whether or not they are actually referenced in the text. This is appropriate
% for the sample file to show the different styles of references, but authors
% most likely will not want to use it.

\end{document}